# Modeling Of Blood Vessel Constriction In 2-D Case Using Molecular Dynamics Method

M. Rendi A. S.[1], Suprijadi[2,*], S. Viridi[1]

[1]*Nuclear Physics and Biophysics Research Division, Faculty of Mathematics and Natural Sciences, Institut Teknologi Bandung, Jl. Ganesha 10, Bandung 40132, Indonesia*

[2]*Theoretical High Energy Physics and Instrumentation, Faculty of Mathematics and Natural Sciences, Institut Teknologi Bandung, Jl. Ganesha 10, Bandung 40132, Indonesia*

*\*supri@fi.itb.ac.id*

**Abstract.** Blood vessel constriction is simulated with particle-based method using a molecular dynamics authoring software known as Molecular Workbench (WM). Blood flow and vessel wall, the only components considered in constructing a blood vessel, are all represented in particle form with interaction potentials: Lennard-Jones potential, push-pull spring potential, and bending spring potential. Influence of medium or blood plasma is accommodated in plasma viscosity through Stokes drag force. It has been observed that pressure $p$ is increased as constriction $c$ is increased. Leakage of blood vessel starts at 80 % constriction, which shows existence of maximum pressure that can be overcome by vessel wall.

## INTRODUCTION

It has been well accepted that in medium-to-large arteries blood can be modeled as a viscous, incompressible Newtonian fluid. Although blood is a suspension of red blood cells, white blood cells, and platelets in plasma, its non-Newtonian nature due to the particular rheology is relevant in small arteries (arterioles) and capillaries where the diameter of the arteries becomes comparable to the size of the cells. In medium-to-large arteries, such as the coronary arteries (medium) and the abdominal aorta (large), the Navier–Stokes equations for an incompressible viscous fluid are considered to be a good model for blood flow. Also, a blood flow is governed by the continuity equation [1].

Observing blood flow from particle-based point of view is interesting and challenging. In this work, Navier-Stokes equation [2], which is common in discussing fluid, is not used but another approach based on particle will be. As the case, constriction of blood vessel is chosen. Blood and its vessel will be represented as particles with particular interactions between them.

## BLOOD VESSEL MODEL

In this work, blood vessel is simplified in 2-d case and considered to consist only of vessel wall and blood flow. These two components are modeled based on particles interactions, which requires three types of interaction between particles: (i) interaction between particles that represents blood flow (blood particles), (ii) interaction between particles that represents vessel wall (wall particles), and (iii) interaction between blood particle and wall particle.

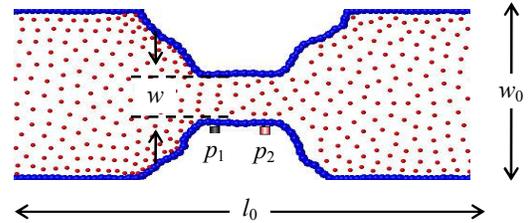

**FIGURE 1**. Model of blood vessel constriction in 2-d case with vessel length $l_0$, vessel width $w_0$, and width of constricted part $w$ equipped with two pressure gauges that measures pressure $p_1$ and $p_2$.

In the proposed model, dispersed red spherical particles represent blood particles and blue spherical particles, that construct chain-like form, represent wall particles as given in Figure 1. All particles are assumed to be spherical.

### Particles interactions

Interaction between blood particle $i$ and $j$ is given through Lennard-Jones potential [3]

$$V_{LJ,ij} = 4\varepsilon\left[\left(\frac{\sigma}{r_{ij}}\right)^{12} - \left(\frac{\sigma}{r_{ij}}\right)^{6}\right], \qquad (1)$$

with $r_{ij}$ is distance between particle $i$ and $j$, $\varepsilon$ represents the strength of the interaction, and $\sigma$ is the equilibrium distance between the two interacting particles [4].

Vessel wall is represented with wall particles that interact with each other through push-pull spring potential $V_s$ and bending spring potential $V_b$ [5]. The first potential governs interaction between wall particle $i$ and $j$ through

$$V_{s,ij} = \frac{1}{2} k_s (r_{ij} - r_0)^2, \quad (2)$$

with $r_0$ is equilibrium distance between two wall particles and $k_s$ is a spring constant. The second potential need three wall particles to govern, particle $i$, $j$, and $k$, through

$$V_{b,ij} = \frac{1}{2} k_b \tan^2 \theta_{ijk}, \quad (3)$$

with $k_b$ is bending constant, and $\theta_{ijk}$ is angle between $r_{ij}$ and $r_{kj}$, which is defined through

$$\cos \theta_{ijk} = \left[ \frac{(\vec{r}_i - \vec{r}_j) \cdot (\vec{r}_k - \vec{r}_j)}{|\vec{r}_i - \vec{r}_j||\vec{r}_k - \vec{r}_j|} \right]. \quad (4)$$

Illustration of potentials in Equation (2) and (3) is given in Figure 2, which is explained the vessel wall in Figure 1.

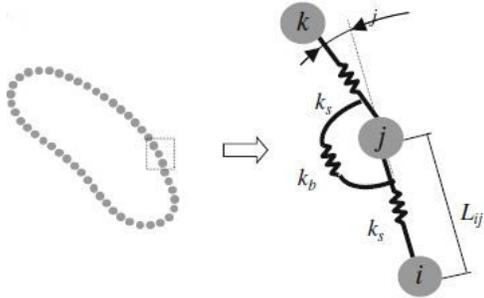

**FIGURE 2.** Model of vessel wall which is consisted of wall particles that interact through push-pull spring potential $V_s$ (with spring constant $k_s$) and bending spring potential $V_b$ (with spring constant $k_b$).

Elastic energy, which is kept during wall vessel deformation due to collision between wall and blood particle, is only summation of Equation (2) and (3) for all wall particles [6].

Blood particles are maintained inside the blood vessel through collision interaction between blood particle and wall particle. The interaction is represented by linear spring-dashpot model [7], which is similar to Equation (2), by neglecting the dissipation part, as long as there is overlap between wall and blood particle.

## Medium influence

It is considered that blood particles move in a fluid or blood plasma that has a certain dynamic viscosity $\mu$ and these particles are suffered a drag force in a form of

$$\vec{F}_{d,i} = -3\pi D_i \mu \vec{v}_i, \quad (5)$$

with $D_i$ is diameter of blood particle $i$.

## Molecular dynamics method

Using conserve principle from gradient theorem,

$$\vec{F} = -\vec{\nabla} V, \quad (6)$$

force forms of potential in Equation (1) – (4) can be obtained. Since all forces are function of position or derivative of position with respect to time $t$, then with Newton second law of motion, dynamics of every particle (blood and wall particles) can be calculated from

$$m_i \frac{d^2 \vec{r}(t)}{dt^2} = \vec{F}_{d,i}(t) + \sum_{j \neq i} \vec{F}_{LJ,ij}(t) + \vec{F}_{s,ij}(t) + \vec{F}_{b,ij}(t), \quad (7)$$

where $m_i$ is mass of particle $i$. Terms in the right side of Equation (7) are from Equation (5) and from Equation (1) – (4) by implementing Equation (6). Through an arbitrary well-known integration method, Equation (7) can be also written a form of

$$\vec{r}_i(t + \Delta t) = A\left[ \frac{d\vec{r}_i(t)}{dt}, ..., \vec{r}_{i-1}(t), \vec{r}_i(t), \vec{r}_{i+1}(t), .. \right], \quad (8)$$

where form of function $A$ depends on the integration method that is chosen to solve Equation (7). In this work Equation (8) is solved numerically using Molecular Workbench (MW), which is an authoring software applying molecular dynamics method [8].

## RESULTS AND DISCUSSION

Table 1 shows parameters used in the simulation designed using MW for model that is previously illustrated in Figure 1.

**TABLE 1.** Simulation parameters.

| Symbol | Value | Unit |
|---|---|---|
| $\rho$ | 1060 | kg/m$^3$ |
| $D$ | 0.5 | µm |
| $\mu$ | 4000 | Pa·s |
| $k_s$ | 0.2 | N/m |
| $k_b$ | 50 | N·m |
| $l_0$ | 50 | µm |
| $w_0$ | 25 | µm |
| $\varepsilon$ | 0.01 | eV |
| $\sigma$ | 0.5 | Å |
| $\Delta t$ | 1 | fs |
| $N$ | 500 | - |

In the simulation number of particles is limited to $N = 500$, which is already including blood and wall particles.

Vessel constriction $c$ is defined as

$$c = 1 - \frac{w}{w_0}, \qquad (9)$$

with $w_0$ is initial width of blood vessel and $w$ is width of middle of vessel, where it is also the constricted part of the blood vessel. Range of values of $c$ is 0 – 95 %. As initial condition a certain value of pressure is applied in a simulation at left side of Figure 1, while the blood flow goes from left to right. Two different values of pressure are used, which are 75 mmHg and 105 mmHg. In order to observe pressure in the constricted part of the blood vessel two pressure gauges are set as previously shown in Figure 1. These pressure gauges give value of pressure $p_1$ and $p_2$, respectively. Average value of these values is calculated and reported as $p$.

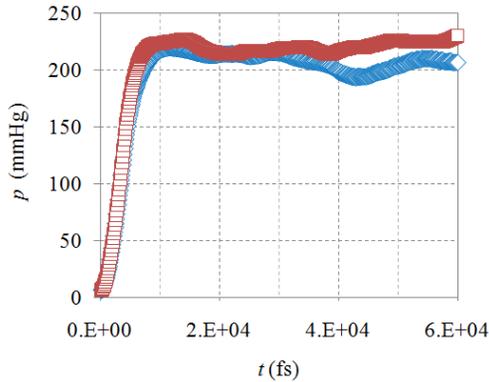

**FIGURE 3**. A typical change of pressure $p$ (averaging from directly measured $p_1$ and $p_2$) in the constricted part of the blood vessel is as function of time $t$ for different initial pressure $p_0$: 75 mmHg (◇) and 105 mmHg (□).

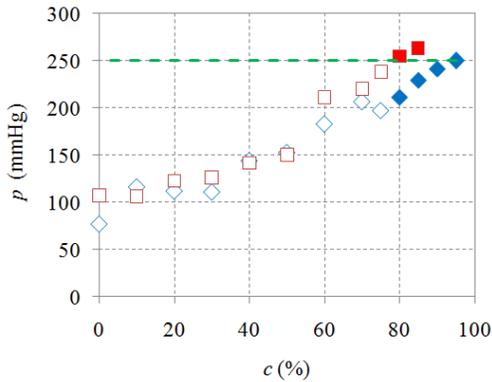

**FIGURE 4**. Measured pressure $p$ in the constricted part of the blood vessel is as function of constriction $c$ for different initial pressure $p_0$: 75 mmHg (◇) and 105 mmHg (□), where solid marker indicates leakage of blood vessel.

A typical change of pressure $p$ as function of time $t$ is shown in Figure 3 for constriction $c$ = 75 %. Pressure $p$ is observed to change from zero to a certain transient value after about 60 ps. This certain value, which is obtained by averaging pressure $p$ after 20 ps, depends on vessel constriction, whose dependence is shown in Figure 4. It can be observed that pressure $p$ is increased as vessel constriction $c$ is increased. At about 80 % of constriction, leakage of blood vessel starts, which is indicated by some blood particles that can penetrate the vessel wall. This phenomenon can be addressed to the pressure inside blood vessel that is too high to be overcome by the vessel wall. Value of 250 mmHg is observed to be the maximum value [9], which is suitable to maximum value about 210 and 250 mmHg for initial pressure 75 and 105 mmHg, respectively, as it can be seen in Figure 4.

## CONCLUSION

A design of computational models of blood vessels constriction in 2-d case has been proposed and tested. It can show existence of maximum blood flow pressure that may induce blood vessel leakage, which is about 210 and 250 mmHg for initial pressure 75 and 105 mmHg, respectively. These values are near to the previously reported one that could be considered as a confirmation to results of this work.